# Reply to Comment on "Φ memristor: Real memristor found" published on arxiv.org/abs/1909.12464

Frank Z. Wang

School of Computing, University of Kent, Canterbury, CT2 7NF, UK
Corresponding author: Frank Z. Wang (frankwang@ieee.org).

**Abstract:** In this reply, we will provide our impersonal, point-to-point responses to the major criticisms (in bold and underlined) in the Comment published on arxiv.org/abs/1909.12464. Firstly, we will identify a number of (imperceptibly hidden) errors in the Comment in understanding/interpreting our physical model. Secondly, we will use a 3$^{rd}$-party experiment carried out in 1961 (plus other 3$^{rd}$-party experiments thereafter) to further support our claim that our invented Φ memristor is memristive in spite of the existence of a parasitic "inductor" effect. Thirdly, we will analyse this parasitic effect mathematically, introduce our work-in-progress (in nanoscale) and point out that this parasitic "inductor" effect should not become a worry since it can be completely removed in the macro-scale devices and safely neglected in the nano-scale devices.

**Re: "However, this property [a typical magnetization curve of ferromagnet/ferrimagnet materials showing that the magnetization switches between two limiting values ($\pm M_s$) in response to a time-dependent (e.g., sinusoidal) input.] is missing in the Eq.(1) model. Therefore, there is a clear disagreement between the response of actual physical ferromagnets/ferrimagnets and the dynamics described by Eq. (1)."**

Our Reply: Thank the Comment authors for their reminder, we also felt it would be a good try to reproduce a typical *m-H* curve from our key equation Eq.(3) [2] (re-written as Eq.(1) in [1] by the Comment authors) in order to test if it features a magnetic core reasonably. Using similar parameters and a typical *M-H* loop used in [1], the reproduced *m-H* curves are shown in Fig.1 (b) and (c), in contrast to the typical m-H curve [3] in (a). As can be seen, our curves are close to the typical one in terms of revealing a hysteresis loop between *m* and *H*.

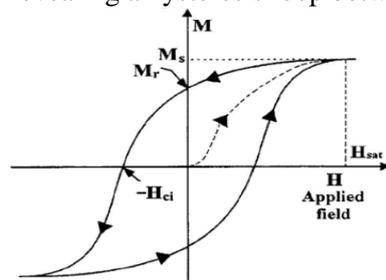

(a) Typical M-H;

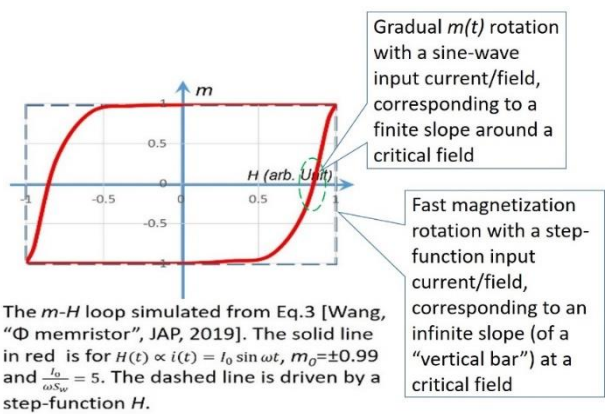

The *m-H* loop simulated from Eq.3 [Wang, "Φ memristor", JAP, 2019]. The solid line in red is for $H(t) \propto i(t) = I_0 \sin \omega t$, $m_0 = \pm 0.99$ and $\frac{I_0}{\omega S_w} = 5$. The dashed line is driven by a step-function $H$.

Gradual *m(t)* rotation with a sine-wave input current/field, corresponding to a finite slope around a critical field

Fast magnetization rotation with a step-function input current/field, corresponding to an infinite slope (of a "vertical bar") at a critical field

(b) Wang's simulated m-H loop based on Eq.3;

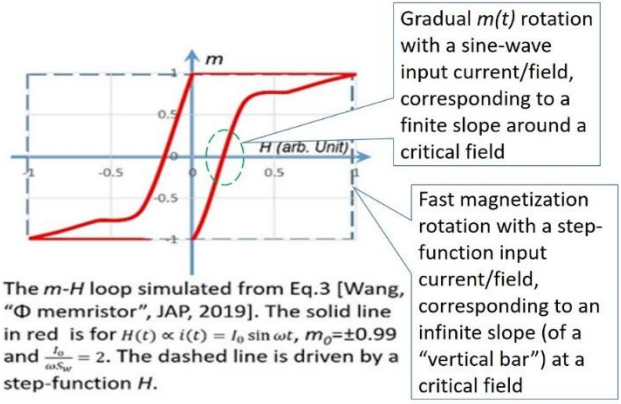

(c) Wang's simulated m-H loop based on Eq.3;

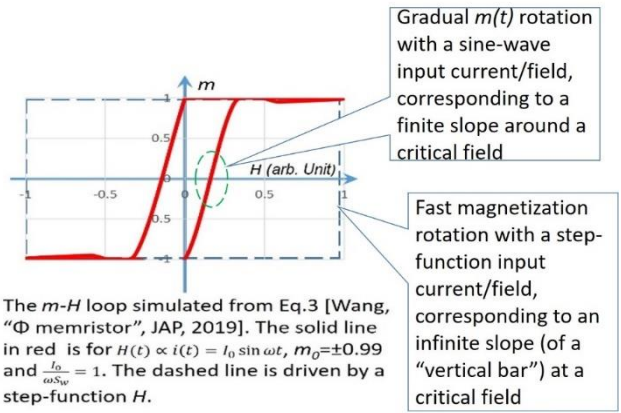

(d) Wang's simulated m-H loop based on Eq.3;

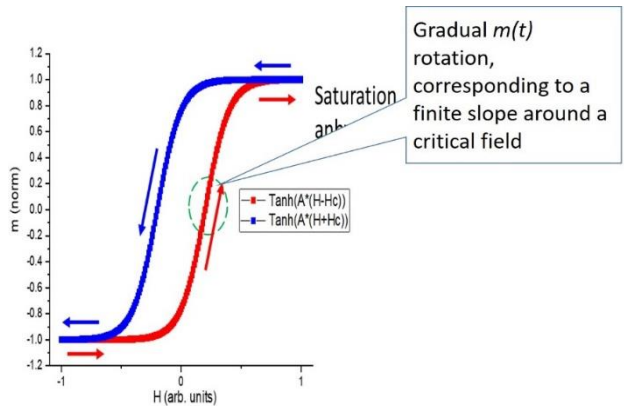

(e) The JAP editors' simulated m-H loop based on m=tanh⁡(A*(H±H_C ))] with a sine-wave input current.

Fig.1 (a) Magnetization, *M*, vs magnetic field, *H*, curve for a ferromagnetic/ferromagnetic material. Reprinted with permission from Ref. [3]. (b-d): Normalized magnetization, *m*, versus *H* plotted using $m(t) = tanh\left[\frac{q(t)}{S_W} + tanh^{-1}m_0\right]$ extracted from Ref. [2] for $H(t) \propto i(t) = I_0 \sin \omega t$ and $q(t) = -\frac{I_0}{\omega}\cos \omega t$. (e) The JAP editors' simulated m-H loop.

Although $H_C$ is not directly seen in my Eq.3, we mentioned "The threshold for the magnetization switching is automatically taken into account because the switching coefficient is defined based on the threshold field $H_0$, which is of one to two times the coercive force $H_C$ [29][30][31]." in our original JAP paper. As shown in Fig.1(e)], our original formula (Eq.3): $m(t) = tanh\left[\frac{q(t)}{S_W} + C\right] = tanh\left[\frac{1}{S_W}(q(t) + S_W tanh^{-1}m_0)\right] = tanh\left[\frac{1}{S_W}(q(t) \pm S_W tanh^{-1}|m_0|)\right]$ is mathematically equivalent to the JAP editors' recommended one: $m =$

tanh($A*(H \pm H_C)$), in terms of "using two Tanh and applying a horizontal shift to each branch to get hysteresis" (in spite of a 90° phase shift of the two branches).

We really don't know why the Comment authors produced a dramatically different curves (Fig.1 of [1]) from ours. Intuitively, when $H(t) = 0$, we should have $\omega t = \cdots, -\pi, 0, \pi, \cdots$, $q(t) = -\frac{I_0}{\omega}\cos \omega t = \cdots, \frac{I_0}{\omega}, -\frac{I_0}{\omega}, \frac{I_0}{\omega}, \cdots$ and $m(t) = tanh\left[\frac{q(t)}{S_W}\right] = \cdots, tanh\left[\frac{I_0}{\omega S_W}\right], -tanh\left[\frac{I_0}{\omega S_W}\right], tanh\left[\frac{I_0}{\omega S_W}\right], \cdots$, which implies $m$ is symmetric about the $H$ axis when $H$ is zero. It seems that our curves are more reasonable.

### Re: "Fig. 5 of Ref. [1] can not be reproduced using the analytical expression for the voltage across the memristor, Eq. (2)..."

Our Reply: After having read this criticism over and over, we eventually realised that the Comment authors may have made some (unfortunately lethal) mistakes in understanding/interpreting our original physical model. For example, they imprudently extended "*I*" to "*I(t)*" in re-writing our original equation (Eq.(3) in [2]) as Eq.(2) in their Comment [1], replaced "*I(t)*" with a sinusoidal function and then complained there was no reproducibility for Fig.5. They must have overlooked our stated condition ("a step-function excitement current is applied", "… approaching constant *I*") that was located just 2-3 lines above the corresponding equation [Eq.(7)] and Fig.4 [in which "*I₁*" and "*I₂*" obviously represent the two (fixed, unchanged) amplitudes of the current in the graph] in our original JAP paper [2]. Throughout our JAP paper [2], the upper-case letter "*I*" was used to denote a constant current, rather than a changing one. Even if one overlooks this important condition ("step-function", "constant *I*") but follows our deduction from Eq.(6) to Eq.(7) [2], one will still realize "*I*" must be a constant otherwise the deduction wouldn't be continued smoothly because $q(t) = I \cdot t$ is established only if "*I*" is a constant [$q(t) = \int i(t)dt$ if the current *i(t)* is not a constant]. By the way, we used the lower-case letter with an explicit expression "*i(t)*" to denote the (changing) current as a function of time. Actually, we had deducted step by step a detailed equation for the sinusoidal response as Eq.(12) (depicted in Fig.5) in our original paper [2], which was obviously overlooked again by the Comment authors because they themselves were trying to work out the sinusoidal response (unnecessarily since we had already done it in [2]) from a wrong equation Eq.(2) [1] (that was originally for a constant input only). In short, in our original JAP paper [2], Eq.(7) is depicted in Fig.4 as the response to a constant input whereas Eq.(12) in Fig.5 as the response to a sinusoidal input. These two scenarios have the different equations to describe them (one can imagine the equation for a constant input should be much simpler than that for a sinusoidal input). In any case, one should not use the constant-input equation to find the sinusoidal response.

### Re: "We continue to argue that it was the correct classification, namely that a current-carrying wire strung through a magnetic core is not a memristor, …"

Our Reply: We found a historic reference as a 3rd-party evidence (listed as Ref. [30] in our original JAP paper [2]) to support our claim that a magnetic core is memristive. In 1961, Cushman characterized magnetic switch cores in order to investigate the flux-charge concept and relate the core characteristics to actual circuit application [4]. His measurements are reprinted in Fig2.

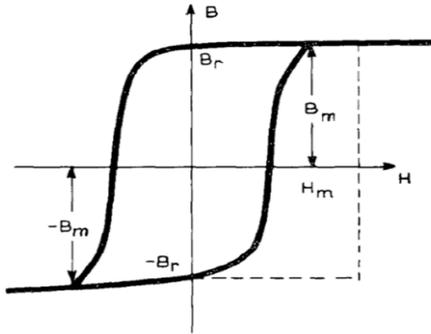
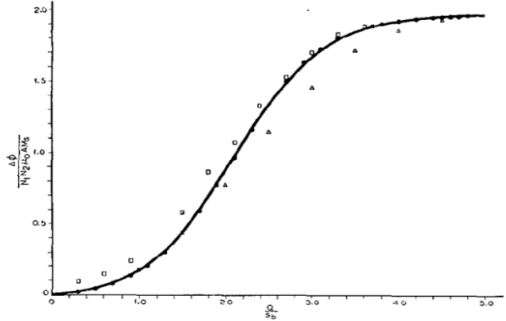

(a) The toroidal magnetic cores have static B-H loops whose shape resembles this;

(b) The measured Φ-Q pairs and a fit Φ-Q curve.

Fig.2 Characterization of magnetic switch cores (reprinted with permission from Ref. [4]). A typical *B-H* loop of ferromagnet/ferrimagnet materials is shown in (a), which further evidences our 1st reply. In (b), it can be seen that the flux change that occurs in a magnetic switch core depends only on the effective charge *Q* that has flowed through the core and the initial state of magnetization. Putting aside its multi-terminal structure, a magnetic core was found to be memristive in terms of satisfying both Chua's three criteria [5] and Georgiou's new three criteria [6] for the ideal memristor.

Putting aside its multi-terminal structure, a magnetic core was found to be memristive in terms of satisfying both Chua's three criteria (1. Nonlinear; 2. Continuously differentiable; 3. Monotonically increasing.) [5] and Georgiou's new three criteria (1. Nonlinear; 2. Continuously differentiable; 3. Strictly monotonically increasing.) [6] for the ideal memristor. Interestingly, this characterization [4] that might have led to the invention of a memristor took place 10 years before Chua originally theorized memristor in 1971 [5]. By the way, our invention was not inspired by this paper [4]. As mentioned in our original JAP paper [2], we followed the 1st principle by putting a conductor (carrying electricity) and a magnet (carrying magnetism) together to (simply, brutally) introduce the magnetism/electricity interaction required in Chua's original definition of memristor [5].

After the above Cushman characterization in 1961 [4], other similar measurements of magnetic cores were carried out [7][8][9]. Fig.3 shows the comparison of the behaviour of a magnetic core anticipated by a charge model [7] and that of our Φ memristor measured experimentally [2]. A reasonable agreement can be seen. It is worth mentioning that our physical model anticipated a behaviour in Fig.4 of Ref. [2], which is similar to Fig.3(a) here (Fig.1 of Ref. [7]).

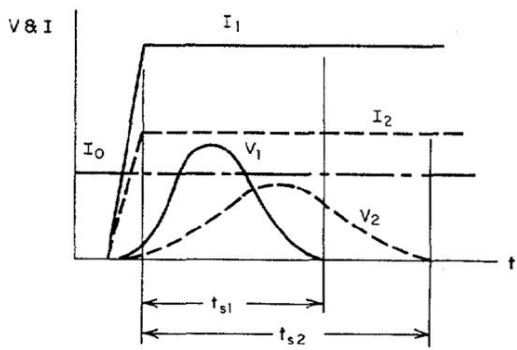
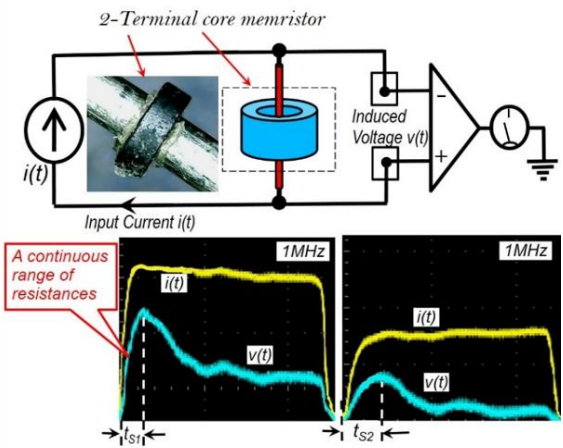

(a) Anticipated behaviour by a charge model of the storage core [7];

(b) Our Φ memristor.

Fig.3 The behaviour comparison of a magnetic core and our Φ memristor (reprinted with permission from Ref. [7] & [2], respectively). Our experiment in (b) exhibits clearly the memristance whose value [$M(q) = \frac{v(t)}{i(t)}$] is a function of charge *q* [the history of the (changing) current *i(t)*], which meets the memristor definition [5].

Note that, as mentioned in [2], our Φ memristor is still a new invention (rather than a simple replica of traditional magnetic cores) with a different structure from that of a standard magnetic core memory cell: ours has two-terminals only whereas a magnetic core has several wires (typically two X/Y lines for coincident-currents plus one sense/inhibit line).

**Re: "a current-carrying wire strung through a magnetic core is … simply an inductor with memory."**
Our Reply: In our original JAP paper [2], we said our Φ memristor is memristive with a parasitic "inductor" effect. It seems that both of us and the Comment authors recognized the co-existence of the memristivity (or memory) and the inductivity effects in this core structure. Therefore, the ultimate argument could only be: which effect is dominant, memristivity or inductivity?

Regarding the parasitic "inductor" effect that concerns the Comment authors seriously, we had already elaborated in details in our original JAP paper [2] as below:

"Note that a parasitic "inductor" effect exists in the core structure, which was observed as sharp transient spikes caused by the sudden change of the input current ($V_L(t) = L\frac{di(t)}{dt} \neq 0$). Fortunately, this effect is narrowly constrained at the rise/fall edges of the step function. We removed this high-frequency noises by simply using the "compensation adjustment" function (a low-pass filter, LPF) of an oscilloscope (GW Instek GDS-1072B). In the most time of the cycle (especially in the memristive region of our interest), no "inductor" exists as $V_L(t) = L\frac{di(t)}{dt} = 0$.

Definitely, the *Φ* memristor is neither "an inductor with memory" nor a mem-inductor although there may exist the parasitic "inductor" effect. Actually, in Section V, the uniqueness of those circuit elements with memory is demonstrated in ANA whose (adaptive) time constant ($\sqrt{L(q(t))C}$) is determined by a mem-inductor (whose inductance is a function of charge) [19]. What makes our *Φ* memristor different from mem-inductor and others is that its resistance is a function of charge (Eq.8 & Fig.8). Experimentally, no capacitive or inductive effects was observed (otherwise there should be a phase shift between the current and voltage) by taking two measures: 1. Using the step function excitement current only; 2. Using a low-pass filter that passes signals with a frequency lower than a certain cut-off frequency and attenuates noises with frequencies higher than the cut-off frequency.

As we tested, no voltage was generated when we applied a current through a bare wire without any magnetic core (since an ideal wire has no resistance). It is the magnetization in the core that induces a voltage through its reversal, which has been observed."

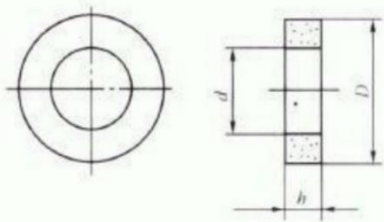

Fig.4 Geometry of a magnetic core for the inductance calculation (reprinted with permission from Ref. [10]).

As a new technical contribution in this reply, this parasitic inductance of a magnetic core (Fig.4) can be described mathematically by the following equation [10]:

$$L = 0.4\pi\mu N^2 \frac{A}{l} \times 10^{-5} (mH), \qquad (1)$$

where *L* is the inductance (*mH*) of the magnetic core, $A = \frac{D-d}{2} \cdot h \ (cm^2)$ is the cross-sectional area of the core, $l = \pi \cdot \frac{D+d}{2}$ *(cm)* is the average length of the core, $\mu$ is the permeability of the core, *N* is the number of the turns in the coil (*N=1* in the Φ memristor), *D (cm)* is the outer diameter of the core, *d (cm)* is the inner diameter of the core, and *h (cm)* is the height of the core.

If we always take a fixed aspect ratio *D=2d=2h,* Eq.(1) can be simplified to:

$$L = 0.4\pi\mu N^2 \frac{h}{3} \times 10^{-5} (mH), \tag{2}$$

which implies the inductance of a magnetic core scales with its physical size. This is really encouraging in the sense that, in principle, the Φ memristor in nanoscale may have a negligible parasitic inductance. Actually, this analysis agreed with what we observed experimentally in a nanoscaled Φ memristor we are now intensively working on. Hopefully, we will report this work shortly.

Note that, as shown in Eq.(2), the parasitic inductance in our Φ memristor is not a function of charge (otherwise it will become a mem-inductor [11]) whereas its resistance (memristance) is a function of charge as below (Eq.(8) in [2]):

$$M(q) = \frac{d\varphi}{dq} = \frac{\mu_0 S M_s}{S_W} sech^2 \left( \frac{q(t)}{S_W} + tanh^{-1} m_0 \right). \tag{3}$$

This equation (and the whole physical model) is experimentally verified by Fig.8-10 in [2]. This is why we said our Φ memristor is memristive with a parasitic "inductor" effect. In theory, our claim is based on the above memristivity and inductivity expressions [Eq.(3) & Eq.(2) in this reply].

**Re: "… the design of the \negative memristor" emulator (Fig. 19 of Ref. [1]), as it stands, is incomplete… it is not obvious why the switching resistors are negative."**
Our reply: In our original JAP paper [2], we mentioned "a locally active memristor", "a negative memristor is active and artefact", "the (negative) memristor has an internal source of power, such as light, chemical or nuclear reactions, or batteries". The negative resistances in Fig.19 of our paper [2] is actually emulated by operational amplifiers. We apologize for not detailing this in the caption.

**Re: "In conclusion, we have demonstrated that the Φ memristor Wang et al. claim to have "discovered" [1] is, in fact, not a memristor. The fundamental error made by Wang et al. can be traced to their use of an incorrect model of magnetization dynamics."**
Our Reply: As elaborated above, the Comment authors have made a number of (imperceptibly hidden but unfortunately lethal) mistakes in understanding/interpreting our physical model, which led them to a wrong conclusion. In this reply, we used a 3[rd]-party experiment carried out in 1961 (plus other 3[rd]-party experiments thereafter) to further support our claim that our invented Φ memristor is memristive in spite of the existence of a parasitic "inductor" effect.

**The Reply Summary**
To be clear, we still welcome criticisms/comments/suggestions from other professionals to our work although it had been peer-reviewed before its publication, which inspired us to re-think deeply and, if possible, improve it further (e.g. we added a new technical contribution in this Reply that is an analysis of the parasitic inductance of our invented Φ memristor to address the Comment authors' concerns). On the other hand, other readers may also benefit from this comment/reply correspondence, especially in this case we may have already surprised the physics/electronics/computer communities with a very unusual observation against conventional wisdom: a historic, extinct magnetic core could be memristive if we just creatively simplified its structure! We reckon it may not be uncommon for other readers to have similar questions/concerns/confusions.

We'd like to reiterate our original claim in our JAP paper: our invented Φ memristor is memristive with a parasitic "inductor" effect. In spite of the existence of this parasitic inductance, one should not deny its memristivity, as if one should not deny the fact that a real-world resistor is resistive in spite of the existence of the (inevitable) parasitic inductance and/or capacitance. This claim is based on our physical model (it is neither "incorrect" nor "erroneous" at all as elaborated above) and experiments (including 3rd-parties' ones). This parasitic inductance should not become a worry since it can be fully removed in the macro-scale devices (with a dominant memristive behaviour if we only apply a constant input signal) and, more remarkably, safely neglected in the nano-scale devices.

At the end of this reply, we stress that the importance of inventing the Φ memristor [2] is to experimentally verify the existence of the real memristor according to its original definition in 1971 [5] and, as a seminal attempt, open a new way to prototype a "real" memristor based on the direct magnetism-electricity interaction. As reported in [2], it is so far a binary memristor with only two stable states due to the usage of a magnetically uniaxial anisotropic material. We had provided a number of schemes [2] to make it more comprehensively functioning. As an exciting work-in-progress, our next-generation Φ memristor will be fully equipped with some remarkable features (nanoscale, multi-stability, no parasitic "inductor", two terminal) but it is obviously beyond the scope of this reply.